\documentclass{pnastwo4arxiv}

\usepackage{amssymb,amsfonts,amsmath}
\usepackage{natbib}
\usepackage{graphicx}
\bibpunct{(}{)}{,}{a}{,}{,}

\begin{document}

\title{About incoherent inference}

\issuedate{\today}
\copyrightyear{2010}
\volume{C.P.~Robert}
\issuenumber{Not submitted}
\footlineauthor{C.P.~Robert}
\author{Christian P.~Robert\affil{1}{Universit\'e Paris-Dauphine, CEREMADE, Paris, France}
}
\contributor{ }

\maketitle

\begin{article}


\abbreviations{ABC: approximate Bayesian Computation}

\dropcap{I}n \cite{templeton:2010},
the Approximate Bayesian Computation (ABC) algorithm \citep[see, e.g.,][]{pritchard:seielstad:perez:feldman:1999,
beaumont:zhang:balding:2002, marjoram:etal:2003,ratmann:andrieu:wiujf:richardson:2009} is criticised on mathematical and 
logical grounds: ``the [Bayesian] inference is mathematically incorrect and formally illogical". 
Since those criticisms turn out to be bearing on Bayesian foundations rather than on the computational methodology
they are primarily directed at, we endeavour to point out in this note the statistical errors and inconsistencies in \cite{templeton:2010}, 
refering to \cite{clade:2010} for a reply that is broader in scope since it also covers the phylogenetic aspects of nested clade versus
a model-based approach.

\section{Coherence}
\cite{templeton:2010} mostly uses arguments found in \cite{templeton:2008} and already answered in \cite{clade:2010}.
However, the tone adopted in this PNAS paper is harsher and has a wider scope than in the earlier paper, in that it
contains a foundational if inappropriate critical perspective on Bayesian model comparison. All of the arguments presented
in Templeton's tribune against the ABC ``method" \citep{tavare:balding:griffith:donnelly:1997} actually aim at exposing the 
incoherence of the Bayesian approach. The major point of contention is that Bayes factors are mathematically incorrect because
they contradict basic logic by being incoherent.
The notion of coherence used in \cite{templeton:2010} is borrowed from \cite{lavine:schervish:1999}.
Those authors introduced this notion to criticise Bayes factors in the limited sense that those may be 
nonmonotonous in the alternative hypothesis---in cases when monotony is relevant---, 
and thus that posterior probabilities---which are coherent---should be used instead in a correct decision theoretic perspective.

\section{Bayes factors}
The core of the Bayesian paradigm is to incorporate all aspects of uncertainty within a prior distribution on the parameter 
space and all aspects of decision consequences within a loss function in order to produce
a single inferential machine that provides the ``optimal" solution \citep{berger:1985}. Posterior probabilities 
and hence Bayes factors \citep{kass:raftery:1995} are the
product of this inferential machine when the goal is the selection of a statistical model. We recall that a 
Bayes factor, of the form
$$
B_{12}^\pi(x) = \dfrac{\int_{\Theta_1} \pi_1(\theta_1) f_1(x|\theta_1)\,\text{d}\theta_1}
{\int_{\Theta_2} \pi_2(\theta_2) f_1(x|\theta_2)\,\text{d}\theta_2} = \dfrac{m_1(x)}{m_2(x)}\,,
$$
compares the marginal likelihoods at the observed data $x$ under both models under comparison. The suggestion of \cite{templeton:2010}
to ``incorporate the sampling error of the observed statistic" is therefore exhibiting a misunderstanding of the above Bayesian 
construction since the posterior distributions naturally incorporate the sampling errors $f_1(x|\theta_1)$ and $f_2(x|\theta_2)$ 
under both models.

Templeton's (2010) first argument against Bayes factors, namely that
"the probability of the nested special case must be less than or equal 
to the probability of the general model within which the special case 
is nested. Any statistic that assigns greater probability to the special case is incoherent", 
proceeds from the ``natural" argument that larger models should have 
larger probabilities by an encompassing analogy. (Note that the notion of defining ``the" probability over the collection of 
models that Templeton seems to take for granted does not make sense outside a Bayesian framework.)
The author presents a Venn diagram to further explain why a larger set should 
have a larger measure, as if this simple-minded analogy was relevant in model choice settings. We found similar arguments
in the recent epistemological book by \cite{sober:2008} as well as in \cite{popper:1959}. This reductive viewpoint does not account
for the fact that in Bayesian model choice, different models induce different parameters spaces and that those parameter spaces 
are endowed with orthogonal measures, especially if those spaces are of different dimensions. When the smaller parameter space
corresponds to the restriction $\theta_1=0$, the measure of this constraint is zero in the larger space, i.e.~$P(\theta_1=0|M_2)=0$,
when the parameter space is continuous. As stressed by \cite{jeffreys:1939}, testing for point null hypotheses (and hence for
nested models) requires a drastic change of dominating measure so that both the null and the alternative hypotheses have a positive
probability. This implies defining versions of the prior distribution over both parameter spaces.
Therefore, talking of nested models having a ``smaller" probability than the encompassing model or of ``partially overlapping models" 
does not make sense from a measure theoretic (hence mathematical) perspective. In other words, the measure of the event is conditional
on the model considered. (The fifty-one occurences of the words {\em coherent} or {\em incoherent} in the paper do not bring additional 
scientific weight to the argument.)

\section{Bayesian model comparison}
When \cite{templeton:2010} calls to logic for rejecting ``incoherent" probability orderings on models, he
rejects the fact that marginal likelihoods are in the same scale and can be added within the denominator
of posterior probabilities, namely
$$
\text{Pr}(M_i|x) = \dfrac{\Pi_i m_i(x)}{\sum_{j=1}^k \Pi_j m_j(x)}\,,
$$
using standard notations \citep{berger:1985,robert:2001}. His argument is that the denominator is proportional to the probability
of the union of several models and hence that the probabilities of the intersections of the overlapping hypotheses [or models] must
be subtracted". Another Venn diagram explains why this basic consequence of Bayes formula is "mathematically and logically incorrect" 
and why marginal likelihoods cannot be added up when models "overlap". According to Templeton, "there can be no universal denominator, 
because a simple sum always violates the constraints of logic when logically overlapping models are tested". Once more, this simply 
shows a poor understanding of the probabilistic modelling involved in model choice: The argument fails because of the measure-theoretic
assumptions separating models and because model choice ultimately involves the selection of one single model, hence the rejection of all
other models. There cannot be a posterior weight on any intersection for this reason.

A second criticism of ABC (i.e. of the Bayesian approach) is that model choice requires a collection of models and cannot decide outside 
this finite and therefore incomplete collection. The very purpose of a Bayesian model choice procedure exactly aims at selecting the
most likely model among all available, rather than rejecting a given model when the data is unlikely. Studies like 
\cite{berger:sellke:1987} have shown the difficulty of reasoning within a single model. Furthermore, \cite{templeton:2010} advocates 
the use of a likelihood ratio test, which necessarily implies using two models with one nested within the other. 

In this paper, Templeton also reiterates the earlier (2008) criticism that marginal likelihoods are not comparable across models, 
because they ``are not adjusted for the dimensionality of the data or the models" (sic!). This point is missing the whole purpose 
of using marginal likelihoods, namely that they account for the dimensionality of the parameter by providing a natural Ockham's razor 
\citep{mackay:2002} penalising the larger model without requiring to specify a dimension penalty. Both BIC and DIC \citep{spiegbestcarl}
are approximations to the exact Bayesian evidence, which shows the intrinsic penalisation thus provided by marginal likelihoods. Note
also that ABC applies the basic principles of a Bayesian model comparison to a summary statistic that is common across models 
\citep{grelaud:marin:robert:rodolphe:tally:2009}, rather than using model specific summary statistics which would then be inconsistent.

\section{Implications of model criticism}
The point corresponding to the quote 
``ABC is used for parameter estimation in addition to hypothesis testing and another source of incoherence is suggested 
from the internal discrepancy between the posterior probabilities generated by ABC and the parameter estimates found by ABC" 
is that, while the posterior probability that $\theta_1=0$ (model $M_1$) is much higher than the posterior 
probability of the opposite (model $M_2$), the Bayes estimate of $\theta_1$ under model $M_2$ is 
``significantly different from zero". Again, this reflects both a misunderstanding of the probability model, namely that 
$\theta_1=0$ is impossible [has measure zero] under model $M_2$, and a confusion between confidence intervals 
(that are model specific) and posterior probabilities (that work across models). The concluding message that ``ABC is a 
deeply flawed Bayesian procedure in which ignorance overwhelms data to create massive incoherence" is thus unsubstantiated.

\section{ABC is only a Monte Carlo scheme}
An issue common to all recent criticisms by 
\cite{templeton:2008,templeton:2010} is the misleading or misled confusion between the ABC method and 
the resulting Bayesian inference. For instance, Templeton distinguishes between the incoherence in the 
ABC model choice procedure from the incoherence in the Bayes factor, when ABC is used as a computational 
device to approximate the Bayes factor. In the current case, the Bayes factor can be directly derived from 
the ABC simulation since the (accepted or rejected) proposed values are simulated from $\pi(\theta)f(x|\theta)$
(modulo a numerical approximation effect). 
This does not turn the Bayes factor into an ABC or simulation-based quantity. There is therefore no inferential 
aspect linked with ABC,  {\em per se}, it is simply a numerical tool to approximate Bayesian procedures and, with 
enough computer power, the approximation can get as precise as one wishes. 

One of the arguments in \cite{templeton:2010} relies on the following representation of the "ABC equation" (sic!)
$$
P(H_i|H,S^*) = \dfrac{G_i(||S_i-S^*||) \Pi_i}{\sum_{j=1}^n G_j(||S_j-S^*||) \Pi_j}
$$
where $S^*$ is the observed summary statistic, $S_i$ is "the vector of expected (simulated) summary statistics under model $i$" 
and "$G_i$ is a goodness-of-fit measure". Templeton states that this "fundamental equation is mathematically incorrect in every 
instance (..) of overlap." This representation of the ABC approximation is again misleading or misled in that the simulation 
algorithm ABC produces an approximation to a posterior sample from $\pi_i(\theta_i|S^*)$. The resulting approximation to the 
marginal likelihood under model $M_i$ is a regular Monte Carlo step that replaces an integral with a weighted sum (an average), not a 
"goodness-of-fit measure" and the $S_i$'s are replicated many times. The subsequent argument  of Templeton's about 
the goodness-of-fit measures being "not adjusted for 
the dimensionality of the data" (re-sic!) and the resulting incoherence is therefore void of substance. The following argument 
repeats an misunderstanding stressed above with the probabilistic model involved in Bayesian model choice: the reasoning that, if
$$
\sum_j \Pi_j = 1
$$
"the constraints of logic are violated [and] the prior probabilities used in the very first step of their Bayesian analysis 
are incoherent", does not assimilate the issue of measures over mutually exclusive spaces.

In \cite{templeton:2010}, ABC is presented as allowing statistical comparisons among simulated models: 
"ABC assigns posterior probabilities to a finite set of simulated a priori models." The simulation aspect
is treated with suspicion and opposed to "standard classical tests", even though the method is simply 
replacing an intractable integral with a convergin average. Once more, there is no statistical flaw that
can be attributed to ABC since this is a purely numerical method. The models under comparison are therefore
the same as those studied by ``standard classical tests" and what is simulated is a sample from the posterior
distribution associated with this model, not the model itself.

\begin{acknowledgments}
This work was partly supported by the Agence Nationale de la Recherche (ANR, 212,
rue de Bercy 75012 Paris) through the 2009 project ANR-08-BLAN-0218 {\sf Big'MC}
and the 2009 project ANR-09-BLAN-01 {\sf EMILE}.
\end{acknowledgments}


\end{article}
\end{document}